\documentclass[prl,amsmath,amssymb,aps,twocolumn]{revtex4-1}

\usepackage{graphicx}% Include figure files
\usepackage{dcolumn}% Align table columns on decimal point
\usepackage{bm}% bold math
\usepackage{epstopdf}
\usepackage{color}

\begin{document}

\preprint{APS/123-QED}

\title{Many-body kinetics of dynamic nuclear polarization by the cross effect}

\author{A. Karabanov, D. Wi\'sniewski, F. Raimondi, I. Lesanovsky, and W. K\"ockenberger}

\affiliation{School of Physics and Astronomy, University of Nottingham, \\University Park, NG7 2RD, Nottingham, UK}
\affiliation{Centre for the Mathematics and Theoretical Physics of Quantum Non-equilibrium Systems,
University of Nottingham, Nottingham NG7 2RD, UK}

\date{\today}% It is always \today, today,

\begin{abstract}
Dynamic nuclear polarization (DNP) is an out-of-equilibrium method for generating non-thermal spin polarization which provides large signal enhancements in modern diagnostic methods based on nuclear magnetic resonance. A particular instance is cross effect DNP, which involves the interaction of two coupled electrons with the nuclear spin ensemble. Here we develop a theory for this important DNP mechanism and show that the non-equilibrium nuclear polarization build-up is effectively driven by three-body incoherent Markovian dissipative processes involving simultaneous state changes of two electrons and one nucleus. Our theoretical approach allows for the first time simulations of the polarization dynamics on an individual spin level for ensembles consisting of hundreds of nuclear spins. The insight obtained by these simulations can be used to find optimal experimental conditions for cross effect DNP and to design tailored radical systems that provide optimal DNP efficiency.
\end{abstract}

\maketitle

\noindent
{\bf\em Introduction} --- Spectroscopy and imaging techniques based on nuclear magnetic resonance (NMR) are used in many important
applications, ranging from materials sciences to biophysics and medical diagnostics. The NMR signal arises from the Zeeman effect, which at thermal equilibrium gives rise to a weak polarization of the nuclear spins. The sensitivity of NMR can be significantly enhanced by dynamic nuclear polarization (DNP), an out-of-equilibrium method that involves the microwave driven transfer of
the much stronger polarization of unpaired electrons to the nuclear spin ensemble via the electron nuclear hyperfine interaction \cite{o-53,ag-82,w-16}. The resulting many-body dynamics is highly intricate and in particular depends on whether the electrons hosted by paramagnetic centres interact or not. In the case that they do not or only weakly interact, the effective mechanism for polarization transfer is the solid effect (SE) \cite{j-57,a-58,sj-65,j-63,b-59,ab-64,h-10}, which can be understood within the framework of a central spin model formed by an isolated electron and its nuclear surrounding \cite{k-15}.
\\
In this work we focus on the far more involved situation in which coupled electron spins interact collectively with the nuclear ensemble, thereby creating a polarized out-of-equilibrium state. We consider the particularly relevant scenario of two dipolar coupled unpaired electrons,
which can be found in biradical molecules \cite{s-06,m-09,h-04}, or when two monoradicals are in close proximity \cite{s-12,h-12}. Here, a collectively enhanced polarization transfer between electrons and the nuclear ensemble can occur via the cross effect (CE) \cite{hf-12,k-63,k-64,h-67,hh-67}. We shed light on the underlying complex microscopic dynamics that is generally governed by an interplay of coherent and incoherent processes and derive efficiency conditions of CE DNP and its interplay with SE DNP. Moreover, we show under which circumstances the CE DNP non-equilibrium dynamics can be efficiently simulated with classical kinetic Monte-Carlo methods. This enables for the first time the investigation of the polarization dynamics of hundreds of interacting nuclei on an individual spin level. Our study paves the way towards a systematic analysis, utilization and optimization of realistic many-body CE DNP and may serve as a guidance for the design of optimal DNP regimes and tailor-made polarising agents (biradicals) \cite{rg-04,rg-06,rg-10,rg-15,le-16,le-17,psp-17}. Furthermore, we expect our insights to be applicable to the non-equilibrium dynamics of nitrogen-vacancies in diamond \cite{mp-16,lp-13}, which is becoming  a popular platform for the implementation of quantum sensing and diagnostics.

\begin{figure*}[t]
\includegraphics[scale=0.35]{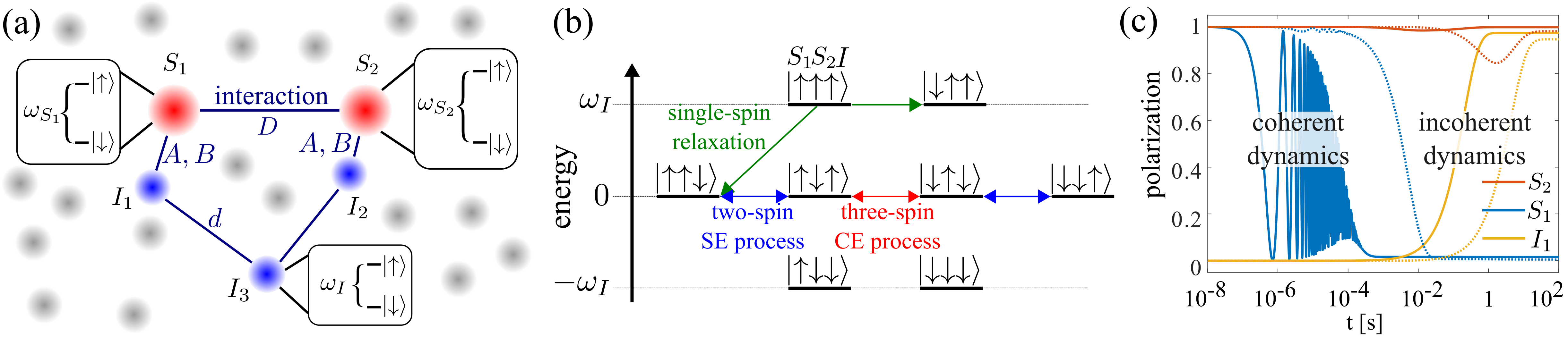}
\vspace{-3mm}
\caption{\label{1}
(a) Schematic representation of the model system (see text for the definition of symbols). (b) Resonance structure of the Zeeman eigenstates and schematics of the basic spin processes. (c) Fast and slow stages of the resonance polarization dynamics of the $n-e_1-e_2$ model simulated with the full master equation with $p=0.034$ ($\omega_S=100$ GHz, $T=70$ K), $\omega_I=145$ MHz, $\omega_1=D=0.5$ MHz, $B=0.1$ MHz, $T_{1e}=3$ ms,  $T_{2e}=50\ \mu$s, $T_{1n}=100$ s, $T_{2n}=1$ ms. Here the efficiency condition (\ref{ec}) is fulfilled with $\eta_1\sim 1$ leading to strongly coherent Rabi oscillations of the first electron (blue curve) in contrast with the incoherent evolution of the nucleus (yellow curve). The dashed lines show a simulation with $p=0.98 \ (T=1$ K), $\omega_1=10$ kHz, $D=100$ kHz, $T_{1e}=1$ s (all other parameters identical to the coherent case),  corresponding to $\eta_1 \sim 100$. In this case only incoherent dynamics is observed. All polarization levels are normalized to the thermal electron polarization $p$.}
\end{figure*}

\begin{figure}[t]
\includegraphics[scale=0.2]{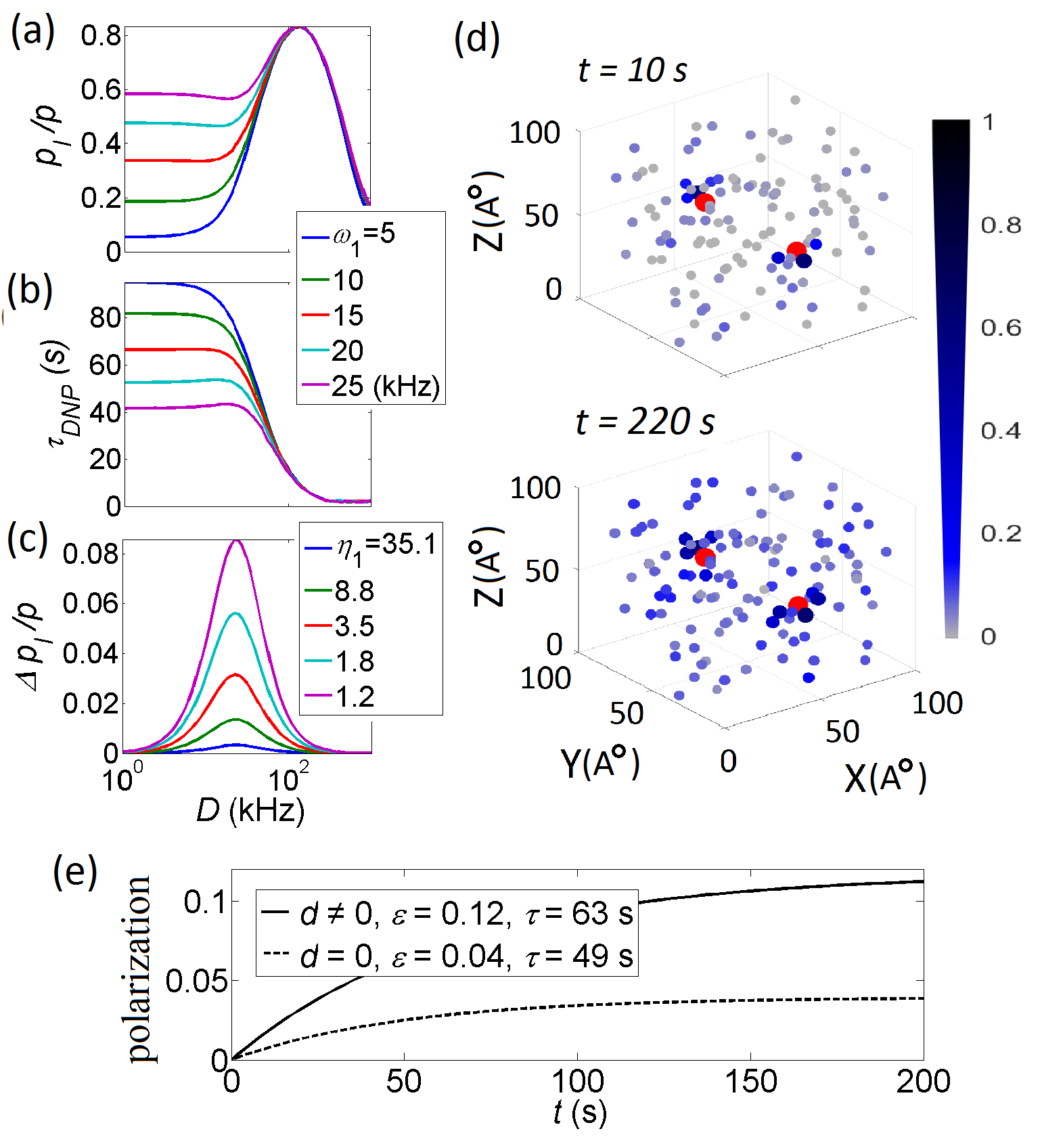}
%\includegraphics[scale=0.17]{F1-Sim.eps}
%\hspace{5mm}
%\includegraphics[scale=0.12]{./Figure222.eps}
%\vskip-5mm
\caption{\label{2}
Steady-state nuclear polarization enhancement (a), build-up time (b) and error between the full master equation and the Zeeman projection (c) as functions of the electron-electron coupling $D$ for different values of the microwave field strength $\omega_1$. In (a,b) calculations are made for the 3-spin $e_1-e_2-n$ model. Calculations in (c) were made with the 5-spin pentagon configuration of FIG.~\ref{1}(a) with $p=0.984$ ($\omega_S=100$ GHz, $T=1$ K), $B_{11}=B_{22}=0.1$ MHz, $d_{13}=d_{23}=10$ Hz (other interaction constants are set to zero), $R_1=1 {\rm s}^{-1}$, $R_2=10^5\ {\rm s}^{-1}$ and other parameters as in FIG.~\ref{1}(c).
(d) Individual polarization build-up of  118 randomly distributed  protons obtained using the Lindblad master equation (\ref{ZP}) and kMC (averaged over 2000 trajectories). The two electrons are shown as red dots. The polarization of the nuclei relative to the thermal polarization of the electron is shown by a grey-blue colour scale and dot size after a short-term (10 s) and long-term (220 s) evolution. (e) Total nuclear polarization build-up for the model described in (d). The effect of spin diffusion can clearly be seen by comparing the build-up in a system with nuclear dipole interaction (solid) to a system in which the dipolar interaction is set to zero (dashed). The enhancement $\epsilon$ and the buildup time $\tau_{\rm DNP}$, $\tau$ in (b,d) are obtained by fitting a monoexponential function for the total nuclear polarization $p_I = \varepsilon(1 - \exp{[-t/\tau])}$.}
\end{figure}

\noindent
{\bf\em Model} ---
The model system that we study is schematically shown in FIG.~\ref{1}(a). It consists of two microwave driven unpaired electron spins ${\mathbf S}_j$, $j=1,\,2$, coupled to a large number of nuclear spins ${\mathbf I}_k$, $k=1,\,2,\,\ldots,\,N$ (assumed to be all spin-$1/2$) at high static field in terms of a Markovian Lindblad master equation for the density matrix $\rho$ in the rotating wave approximation: $\dot\rho=-i[H,\rho]+{\mathcal D}\rho$. The Hamiltonian $H=H_Z+H_{MW}+H_{int}$ describes the Zeeman splitting $H_Z=\omega_II_z+\sum_j\Delta_jS_{jz}$ ($I_z=\sum_kI_{kz}$, $\Delta_j$ are the offsets of the electron Larmor frequencies from the microwave frequency and $\omega_I$ is the nuclear Larmor frequency), microwave irradiation $H_{MW}=\omega_1\sum_jS_{jx}$ (with the strength of the microwave field $\omega_1$) and electron, nuclear and electron-nuclear spin interactions $H_{int}=H_{SS}+H_{II}+H_{SI}$. The electron and nuclear interactions are represented by dipole-dipole secular terms: $H_{SS}=D(3S_{1z}S_{2z}-{\mathbf S}_1\cdot{\mathbf S}_2)$, with the electron-electron coupling strength $D$ and $H_{II}=\sum_{k<k'}d_{kk'}(3I_{kz}I_{k'z}-{\mathbf I}_k\cdot{\mathbf I}_{k'})$ with the nuclear interaction strengths $d_{kk'}$. The electron-nuclear interactions are described by secular and semi-secular parts: $H_{SI}=\sum_{k,j}(A_{jk}I_{kz}+B_{jk}I_{k+}/2+B_{jk}^*I_{k-}/2)S_{jz}$ with the respective interaction strengths $A_{jk}$ and $B_{jk}$. Relaxation is represented by a single-spin Lindblad dissipator ${\mathcal D}={\mathcal D}_S+{\mathcal D}_I$, accounting for electron and nuclear contributions,
${\mathcal D}_S=\sum_j[\Gamma_{1+}{\mathcal L}(S_{j+})+\Gamma_{1-}{\mathcal L}(S_{j-})+
\Gamma_2{\mathcal L}(S_{jz})]$,
${\mathcal D}_I=\sum_k[\gamma_1({\mathcal L}(I_{k+})+{\mathcal L}(I_{k-}))+
\gamma_2{\mathcal L}(I_{kz})]$,
${\mathcal L}(X)\rho\equiv X\rho X^\dagger-(X^\dagger X\rho+\rho X^\dagger X)/2$,
with $\Gamma_{1\pm}=(1\mp p)R_1/2$, $\Gamma_2=2R_2$, $\gamma_1=r_1/2$, $\gamma_2=2r_2$. Here $R_{1,2}$, $r_{1,2}$ are the electron and nuclear longitudinal and transversal relaxation rates, respectively. The parameter $p={\rm tanh}(\hbar\omega_S/2k_BT)\in(0,1)$ is the electron thermal polarization that depends on the average electron Larmor frequency $\omega_S$ and temperature $T$.  

\noindent
{\bf\em Conditions for efficient CE DNP} ---
A fundamental prerequisite for CE DNP is that the Larmor frequencies of the two electrons are separated by the nuclear Larmor frequency, $\Delta_2-\Delta_1\sim\omega_I$. Under this condition the first electron is saturated more efficiently than the second and the arising polarization difference between the two is transferred to a coupled nuclear spin by a three-spin process. In the following we assume that the offsets are chosen such that the matching condition $\Delta_1=\Delta_2-\omega_I=0$ is fulfilled and the polarization difference between the electrons is maximal. In a typical high-field DNP experiment $\vert D\vert,\,\vert B_{jk}\vert\ll\vert\omega_I\vert$ and the electron saturation predominately depends on the microwave term and Zeeman orders of the spin couplings. The projections of the Hamiltonian to the subspaces of the electron spins become $H_j=\omega_1S_{jx}+\bar\Delta_jS_{jz}$, $j=1,\,2$. Here $\bar\Delta_j$ are the effective (operator-valued) electron spin offsets $\bar\Delta_1=2DS_{2z}+\sum_kA_{1k}I_{kz}$, $\bar\Delta_2=\omega_I+2DS_{1z}+\sum_kA_{2k}I_{kz}$ which take on discrete values depending on the up and down orientations of spins. Using the effective Bloch equations for the electron spins, it can be shown [see Supplementary Material A (SM A)] that under the  following condition (assumed to be fulfilled for all discrete values of $\bar\Delta_j$)
\begin{equation}
\eta_1\ll\eta\ll\eta_2,\
\eta_j=(R_2^2+\bar\Delta_j^2)/\omega_1^2,\
\eta=R_2/R_1,
\label{ec}
\end{equation}
the first electron is almost fully saturated, $p_1\ll p$, while the second electron is approximately thermal, $p_2\sim p$, and a  maximal polarization difference between the electrons is generated. Thus, condition (\ref{ec}) defines a parameter regime in which CE DNP is  efficient. 
\\
\noindent
{\bf\em Polarization of single nucleus} ---
In order to get a first idea of the timescales and the role of coherent and incoherent processes in CE DNP, we consider the simplest case of two electrons and a single nucleus coupled only to the first electron via the semi-secular interaction of strength $B$. For simplicity we ignore the secular part of the electron-nuclear interaction $A$ (we refer to this as the $n-e_1-e_2$ model, see  \cite{hf-12} for a detailed discussion). \\
At the resonance $\Delta_1=\Delta_2-\omega_I=0$, the 8-level Zeeman Hamiltonian $H_Z=\omega_I(S_{2z}+I_z)$ in the microwave rotating frame has three sets of degenerate eigenstates with energies $0,\,\pm\,\omega_I$ [see FIG.~\ref{1}(b)]. The density matrix at thermal equilibrium is well approximated by $\rho_{th}=\prod_j(1-2pS_{jz})/8$, with the thermal polarizations of the electrons and nucleus being $p_1=p_2=p$, $p_I=0$, respectively. Proceeding to the full master equation, the non-Zeeman terms of the Hamiltonian connect the Zeeman eigenstates and cause a population exchange between them. The microwave term $\omega_1S_{1x}$ (responsible for the saturation of the first electron) connects states with the same energy. The microwave term $\omega_1S_{2x}$, the flip-flop part of the electron-electron coupling $-D(S_{1+}S_{2-}+S_{1-}S_{2+})/2$ and the semi-secular part of the electron-nuclear interaction $(B_{jk}I_{k+}+B_{jk}^*I_{k-})S_{jz}/2$ (mediating the nuclear polarization build-up) connect states with different energies. Due to $\vert D\vert,\,\vert B\vert\ll\vert\omega_I\vert$, the exchange within the degenerate manifolds occurs on a much faster time-scale than the exchange between the degenerate manifolds. Therefore, the polarization dynamics starting from the thermal state can be divided into two stages: the {\it fast stage} of the first electron saturation and the {\it slow stage} of the nuclear polarization build-up. 
\\
In the fast stage, the polarization of the first electron becomes zero $p_1=0$ while polarization levels of the second electron and nucleus remain unchanged $p_2=p$, $p_I=0$. For small or intermediate values of $\eta_1$, the fast saturation of the first electron is accompanied by coherent Rabi oscillations, resulting in a strong correlation between the observable longitudinal polarization dynamics $\langle S_{1z}\rangle$ and the dynamics in the transversal plane $\langle S_{1x,y}\rangle$ [see solid blue curve in FIG. \ref{1}(c)]. On the contrary, for $\eta_1\gg 1$  the microwave irradiation does not cause Rabi oscillations and the Bloch vector of the first electron remains parallel to the static field while displaying an incoherent longitudinal polarization dynamics $\langle S_{1z}\rangle$ [dashed lines in FIG. \ref{1}(c)]. We will return to this important distinction further below (see also SM A). During the slow stage, the electron-electron flip-flops are energetically matched with the nuclear flips, and effective triple electron-electron-nuclear flips occur that tend to equilibrate the populations of the states $\vert\hspace{-1.2mm}\downarrow\uparrow\downarrow\rangle$, $\vert\hspace{-1.2mm}\uparrow\downarrow\uparrow\rangle$ \cite{hf-12,k-63,k-64,h-67,hh-67}. As a consequence, the thermal polarization of the first electron, which is saturated during the fast stage, is fully transferred to the coupled nucleus [solid yellow curve in FIG.\ref{1}(c)].
\\
The efficiency of CE DNP compared to SE DNP is better because, particularly at low temperatures and weak microwave fields, the nuclear polarization build-up time is much shorter and the nuclear polarization enhancement is much higher. To illustrate this, we consider again the simplest three-spin case but now with a nucleus coupled only to the second electron (the $e_1-e_2-n$ model). Unlike in the $n-e_1-e_2$ model, where CE DNP is the exclusive polarization transfer mechanism, in the $e_1-e_2-n$ model both CE and SE DNP mechanisms can simultaneously participate in the dynamics \cite{hf-12}. In FIG.~\ref{2}(a,b), typical plots of the steady-state nuclear polarization enhancement and build-up time are shown as a function of the electron-electron coupling strength $D$ for different microwave field strengths $\omega_1$. For small $\omega_1$, the steady-state polarization is an order of magnitude higher and the build-up an order of magnitude faster for  the CE case  ($D \approx 100$ kHz) compared to the pure SE case ($D=0$). These differences decrease with increasing $\omega_1$. 

\noindent
{\bf\em Representation by incoherent Markovian dynamics} --- In order to gain insight to the realistic many-body CE DNP dynamics with, e.g., the goal to optimize the physical system parameters and spin geometries, one needs to consider large nuclear ensembles. In a previous work we have shown for the case of SE DNP that it is possible to reduce the quantum master equation dynamics to a set of rate equations that can be efficiently simulated \cite{k-15}. 
\\
Such a strategy is also possible for CE DNP, provided that the saturation of the first electron dominating the fast stage dynamics is incoherent. Generalizing the projective adiabatic elimination technique detailed in \cite{k-15}, we obtain a Lindbladian master equation for the density operator $\rho_Z$, which depends on single, two-body and three-body spin flip processes: 
\begin{equation}
\dot\rho_Z=\bar{\mathcal D}\rho_Z,\quad
\bar{\mathcal D}={\mathcal D}_1+{\mathcal D}_2+{\mathcal D}_3.
\label{ZP}
\end{equation}
Here the single-spin dissipator
$$
\begin{array}{c}
{\mathcal D}_1=\sum_{j=1}^2\left[\Gamma^S_{j+}{\mathcal L}(S_{j+})+\Gamma^S_{j-}{\mathcal L}(S_{j-})\right]+\\[2mm]
\sum_{k=1}^N\Gamma^I_k{\mathcal L}(I_{k+}+I_{k-})
\end{array}
$$
and the rates $\Gamma^{S}_{j\pm}$, $\Gamma_k^I$ arise from spin relaxation, microwave saturation of the electrons and the effective semi-secular saturation of the nuclei. In the double-spin dissipator
$$
\begin{array}{c}
{\mathcal D}_2=\Gamma^{SS}{\mathcal L}(K+K^\dagger)+
\sum_{k=1}^N\Gamma^{SI}_k{\mathcal L}(Y_k+Y_k^\dagger)+\\[2mm]
\sum_{k<k'}\Gamma^{II}_{kk'}{\mathcal L}(X_{kk'}+X_{kk'}^\dagger),\\[2mm]
K=S_{1+}S_{2-},\quad
X_{kk'}=I_{k+}I_{k'-},\quad
Y_k=S_{2-}I_{k+},
\end{array}
$$
the rates $\Gamma^{SS}$, $\Gamma_{kk'}^{II}$ characterize the inter-electron and internuclear flip-flops caused by the dipolar interactions within the electron and nuclear ensembles and the rates $\Gamma^{SI}_k$ describe the effective electron-nuclear flip-flops due to the SE resonance of the second electron and the nuclei. The triple-spin dissipator
$$
\begin{array}{c}
{\mathcal D}_3=\sum_{k=1}^N\Gamma^{SSI}_k{\mathcal L}(Z_k+Z_k^\dagger),\quad
Z_k=S_{1+}S_{2-}I_{k+},
\end{array}
$$
and the rates $\Gamma^{SSI}_k$ represent the effective CE electron-electron-nuclear flips. [For full details and explicit expressions for the Lindbladian rates see SM B].
\\
In the following we analyze in more detail the conditions under which this effectively classical description of CE DNP is applicable. As described in the previous section and illustrated in FIG.~\ref{1}(c), to exclude the strongly coherent saturation dynamics of the first electron, we require $\eta_1\gg 1$, which guarantees the absence of Rabi oscillations (SM A). Combined with condition (\ref{ec}) this leads to a triple inequality that defines a balance between the system parameters for which CE DNP is efficient and the saturation of the first electron is incoherent. In particular, it implies $\eta=R_2/R_1\gg 1$, which is realized in a low-temperature regime. Since the square of the microwave field strength $\omega_1$ is in the denominator of $\eta_1$, condition $\eta_1\gg 1$ also implies a weak microwave field. To illustrate the importance of this condition, we considered a 5-spin system consisting of two electrons ${\mathbf S}_{1,2}$ and three nuclei ${\mathbf I}_{1-3}$ arranged in a pentagon configuration that represents two ``core'' nuclei in close vicinity to the electrons and a ``bulk'' nucleus remote from the electrons, c.f. FIG.~\ref{1}(a), with a symmetric set of interaction strengths. For this small representative system, the exact numerical solution of the full master equation can be compared with the Zeeman projection. The result is shown in FIG.~\ref{2}(c) where we plotted the maximal error $\Delta p_I/p$, $\Delta p_I\equiv\max_{t,k}\big\vert 
p^{\rm FE}_{I,k}(t)-p^{\rm ZP}_{I,k}(t)\big\vert$ (normalized to the thermal electron polarization), over the nuclear polarization build-up between the full master equation and the Zeeman projection as a function of the electron-electron coupling $D$ for different values of the microwave field strength $\omega_1$. It is evident that the error is small for large $\eta_1$ (calculated for values of $D$ at the error peaks) and increases with decreasing $\eta_1$ (still remaining $<10\%$ for $\omega_1<30$ kHz). Note that the maximal build-up error is obtained at the steady-state, and in the case $D=0$ the dynamics of the first electron is decoupled and does not influence the error. 
\\
In contrast to SE DNP, there are electron-electron and three-body electron-electron-nuclear jumps in the effective Lindbladian (\ref{ZP}) of CE DNP. The corresponding rates are given by $\Gamma^{SS}=D^2R_2/2\omega_I^2$, $\Gamma^{SSI}_k=
D^2\vert B_{1k}-B_{2k}\vert^2\tau_{3k}/\omega_I^2$, where the (operator valued) magnitudes $\tau_{3k}$ have dimension of time and depend on the spin transversal relaxation rates $r_2,\,R_2$ and secular electron-nuclear interaction strengths $A_{jk}$ (see SM B for the full expressions). For a good approximation of the polarization dynamics, the condition $\Gamma^{SS},\,\Gamma^{SSI}_k\ll R_2$ must be fulfilled that guarantees that the relevant spin jumps are represented by incoherent Markovian processes (see SM B for details). The other single- and double-spin effective rates in ${\mathcal D}_{1,2}$ are similar to those described in the SE case \cite{k-15}. 
\\
Comparing the CE triple-spin rates $\Gamma^{SSI}_k$ with the SE double-spin rates $\Gamma^{SI}_k$ (given in SM B), we see that for $D^2\gg\omega_1^2$  the CE dominates over the SE while for $D\sim\omega_1$ the CE and SE mechanisms (indicated by the blue and red arrows in FIG.~\ref{1}(b)) equally influence the dynamics (see FIG.~\ref{2}(a,b) as an illustration). Note that condition (\ref{ec}) is violated for large values of $D$,  making the DNP process inefficient.
\\
Where the condition $\eta_1\gg 1$ is not fulfilled, the coherent oscillations in Fig. 1(c), which are not captured by the rate equations (\ref{ZP}), become important. We conjecture, that in this case it is possible to use the state after the fast coherent oscillation as an initial state and an alternative incoherent Lindblad equation similar to Eq.~(\ref{ZP}) can be found to describe the evolution of the system. 

\noindent
{\bf\em Large-scale simulations} ---
To simulate polarization dynamics of large-scale nuclear ensembles, a kinetic Monte Carlo (kMC) scheme can be applied to the incoherent Lindblad master equation (\ref{ZP}) similar to that described for SE DNP \cite{k-15} (see also SM for more details). One such simulation is represented in FIG.~\ref{2}(d,e), showing individual polarization build-up for 118 protons $^1H$ in a random spatial distribution coupled to two electrons. The ideal condition of the electron frequency offsets was used and the parameters for the spin ensemble were chosen to fulfil conditions (\ref{ec}) and $\eta_1\gg 1$. In FIG.~2(e) the two cases of interacting ($d\not=0$) and non-interacting ($d=0$) protons are compared to demonstrate the crucial role of nuclear spin diffusion in the build-up of spin polarization within the nuclear ensemble. The distribution of spin polarization by nuclear spin diffusion simulated in FIG.2 (d,e) is a collective many-body effect of the spin system. Particularly in the case of many polarization sources a large number of nuclei must be considered to obtain a meaningful representation and to analyze conditions under which different DNP mechanisms operate in parallel \cite{s-12}. This was impossible with previous theoretical approaches. Our model allows the CE DNP efficiency to be calculated taking into account the spin geometry of radical molecules embedded in a glass forming matrix. Therefore it provides an important step towards a more rational and less empirical design of radical compounds that have optimized DNP efficiencies.

\newpage
\section{Supplementary Material}
\section{A. Single-spin microwave-driven dynamics}

The microwave-driven single-spin master equation has the form (in the microwave rotating frame and notations similar to those in Eq.~(1) of the main text)
$$
\dot\rho=-i[H,\rho]+\mathcal{D}\rho
$$
with
$$
\begin{array}{c}
H=\bar\omega_1S_x+\bar\Delta S_z,\\[2mm]
{\displaystyle\mathcal{D}=\frac{\bar R_1}{2}\left[(1-\bar p){\mathcal L}(S_{+})+(1+\bar p){\mathcal L}(S_{-})\right]+
2\bar R_2{\mathcal L}(S_{z}).}
\end{array}
$$
In terms of the relative polarization components
$$
\rho=1/2-\bar p\left(XS_x+YS_y+ZS_z\right),
$$
we come to the Bloch equations (for $\bar R_2\gg\bar R_1$) 
\begin{equation}
\begin{array}{c}
\dot X=-\bar\Delta Y-\bar R_2X,\quad
\dot Y=\bar\Delta X-\bar\omega_1Z-\bar R_2Y,\\[2mm]
\dot Z=\bar\omega_1Y+\bar R_1(1-Z).
\end{array}
\label{be}
\tag{S1}
\end{equation}

The steady-state solution where the right-hand sides are all zero is unique and calculated as
$$
\begin{array}{c}
{\displaystyle X=\frac{\bar\Delta Z}{\bar\omega_1\eta'},\quad
Y=-\frac{\bar R_2Z}{\bar\omega_1\eta'},\quad
Z=\left(1+\frac{\eta}{\eta'}\right)^{-1},}\\[5mm]
{\displaystyle\eta'=\frac{\bar R_2^2+\bar\Delta^2}{\bar\omega_1^2},\quad
\eta=\frac{\bar R_2}{\bar R_1}.}
\end{array}
$$
The ratio $\epsilon=\eta/\eta'$ characterizes saturation of the spin: we have a full saturation $Z\sim 0$ for $\epsilon\gg 1$ and no saturation $Z\sim 1$ if $\epsilon\ll 1$. Hence, proceeding from the notations $\bar\Delta,\,\eta',\,Z$ to the notations $\bar\Delta_j,\,\eta_j,\,p_j/p$ of the main text, we come to Eq.~(1) which guarantees the full saturation of the first electron and unchanged polarisation of the second electron, i.e., the maximal polarisation difference between the electrons. 

According to the Bloch equations (\ref{be}), the spin dynamics is composed from the longitudinal dynamics along the $Z$-axis and the transversal dynamics in the plane $(XY)$. In the absence of the microwave $\bar\omega_1=0$, the two dynamics are fully separated and start correlate switching the microwave on $\bar\omega_1\not=0$. The inner transversal dynamics is defined by the equations
$$
\dot X=-\bar\Delta Y-\bar R_2X,\quad
\dot Y=\bar\Delta X-\bar R_2Y
$$  
describing the spin precession accompanied by a $T_2$-decay with the eigenvalues $\lambda_\pm=-\bar R_2\pm i\bar\Delta$. The inner longitudinal dynamics is due to the equation $\dot Z=\bar R_1(1-Z)$ that describes a $T_1$-decay with the rate $\bar R_1$. The rate of exchange between the longitudinal and transversal dynamics is $\bar\omega_1$, so under the condition
\begin{equation}
\vert\lambda_\pm\vert^2=\bar R_2^2+\bar\Delta^2\gg\bar\omega_1^2,\,\bar R_1^2
\label{c}
\tag{S2}
\end{equation}
the transversal dynamics is adiabatically eliminated (see the next section). Thus, the longitudinal dynamics starting from the thermal state $X=Y=0$, $Z=1$ is well described by the projection
\begin{equation}
\dot Z=\bar R_1-\left(\bar R_1+\frac{\bar R_2}{\eta'}\right)Z
\label{Z}
\tag{S3}
\end{equation}
obtained from the third Bloch equation (\ref{be}) after a substitution of the steady-state value $Y=-\bar R_2Z/\bar\omega_1\eta'$. The projection describes then an exponential decay to the steady-state with the effective rate $\bar R_1+\bar R_2/\eta'$. Since $\bar R_2\gg \bar R_1$, condition (\ref{c}) is rediced to the condition $\eta'\gg 1$. In this case the transversal dynamics that satisfies the estimate
$$
X^2+Y^2\sim Z^2/\eta'
$$ 
is negligibly small compared with the longitudinal dynamics. The dynamics starting from the thermal state is almost purely incoherent, i.e., does not appreciably correlate with the transversal dynamics. It is seen that equation (\ref{Z}) is equivalent to the incoherent Lindblad equation
\begin{equation}
\begin{array}{c}
\dot\rho'={\mathcal D}'\rho',\quad
{\mathcal D}'=\Gamma_+{\mathcal L}(S_+)+\Gamma_-{\mathcal L}(S_-),\\[2mm]
{\displaystyle\Gamma_\pm=\frac{\bar R_1(1\mp\bar p)}{2}+\frac{\bar R_2}{2\eta'}}
\end{array}
\label{tr}
\tag{S4}
\end{equation}
in the longitudinal subspace. The second term in $\Gamma_\pm$ is a correction to the effective incoherent rate caused by the presence of the microwave.  

Proceeding from the notation $\eta'$ to the notation $\eta_1$ of the main text, we see that the condition $\eta_1\gg 1$ guarantees an incoherent character of the first electron saturation and applicability of the adiabatic elimination method. In contrast, for small and mediate values of $\eta'$, condition (\ref{c}) does not hold and the transversal dynamics is not adiabatically eliminated. The longitudinal spin dynamics starting from the thermal state strongly correlates with the transversal dynamics featuring strong oscillations in the $(YZ)$-plane called {\it Rabi oscillations}, in full agreement with the dynamics of the first electron considered in the main text and illustrated in FIG.~1(a).

\section{B. Adiabatic elimination procedure}

\subsection{General description}

In order to describe the adiabatic elimination procedure we used to derive the incoherent Lundbladian Eq.~(2), consider first a general linear system (with constant coefficients) defined in a space represented by the direct sum of two subspaces $X_1+X_2$:
\begin{equation}
\dot X_1=L_{11}X_1+L_{12}X_2,\quad
\dot X_2=L_{22}X_2+L_{21}X_1. 
\label{ie}
\tag{S5}
\end{equation}  
Here the operators $L_{11}$, $L_{22}$ describe the inner dynamics in the subspaces $X_1$, $X_2$ respectively. The operators $L_{12}$, $L_{21}$ characterise a dynamic exchange between these subspaces. As justified in the Supplementary Material to our previous work \cite{k-15}, if the total dynamics starts in the subspace $X_1$ that is $X_2(0)=0$, and the inner dynamics in the subspace $X_2$ is much faster than the exchange between the subspaces $X_1$, $X_2$, i.e.,
\begin{equation}
\Vert L_{22}\Vert^2\gg\Vert L_{12}\Vert\cdot\Vert L_{21}\Vert,
\label{ineq}
\tag{S6}
\end{equation}     
then the projection of the total dynamics onto the subspace $X_1$ is well described by the equation
\begin{equation}
\dot X_1=\left(L_{11}-L_{12}L_{22}^{-1}L_{21}\right)X_1
\label{ae}
\tag{S7}
\end{equation}
closed in $X_1$. Equation (\ref{ae}) is obtained by the substitution of the quasi-equilibrium of the second equation 
$$
\dot X_2=0:\quad
X_2=-L_{22}^{-1}L_{21}X_1
$$  
into the first equation of (\ref{ie}). It is assumed additionally that the inner dynamics in the subspace $X_2$ is stable, i.e., the real parts of the eigenvalues of the operator $L_{22}$ are all negative. 

The procedure from inequality (\ref{ineq}) to the projection (\ref{ae}) is called {\it adiabatic elimination} of the subspace $X_2$. We have already seen one of examples of this procedure in the previous section where we proceeded from inequality (\ref{c}) to the projection (\ref{Z}).   

\subsection{Projection to the zero-quantum subspace}  

As an intermediate step towards the final Eq.~(2), we first project the initial master equation $\dot\rho=-i[H,\rho]+{\mathcal D}\rho$ onto the zero-quantum subspace with respect to the resonance Zeeman Hamiltonian $H_Z=\omega_I(S_{2z}+\sum_kI_{kz})$. In other words, we assume first that the subspaces $X_1$, $X_2$ of the adiabatic elimination procedure described in the previous subsection are the subspace of operators commuting with $H_Z$ and the complementary subspace respectively: $[H_Z,X_1]=0$, $[H_Z,X_2]\not=0$. Nonzero eigenvalues of the commutation superoperator $[H_Z,\cdot]$ are multiples of $\omega_I$. Because of the inequality     
$$
\vert\omega_I\vert\gg\vert D\vert,\,\vert B_{jk}\vert,\,\vert A_{jk}\vert,\,\vert d_{kk'}\vert,\,\vert\omega_1\vert,\,r_{1,2},\,R_{1,2}
$$
playing the role of inequality (\ref{ineq}) and typical for high-field low-temperature DNP experimental settings, the subspace $X_2$ is adiabatically eliminated. The projection $\rho_0$ of the total dynamics starting from the thermal state $\rho_{th}\in X_1$ onto the zero-quantum subspace is well described by equation (\ref{ae}) that takes the form
$$
\dot\rho_0=-i[H_0,\rho_0]+{\mathcal D}_0\rho_0.
$$   
Here the explicit forms of the inner operator $L_{11}$ and exchange operators $L_{12}$, $L_{21}$ are found directly from the initial master equation. The inversion $L_{22}^{-1}$ is found as a series in inverse powers of $\omega_I$ using the Krylov-Bogolyubov averaging procedure detailed in Ref.~\cite{k-12,kk-12}. This gives 
$$
\begin{array}{c}
{\displaystyle H_0=\sum_{k<k'}d_{kk'}\left(2I_{kz}I_{k'z}-\frac{X_{kk'}+X_{kk'}^\dagger}{2}\right)+}\\[5mm]
{\displaystyle+\frac{\omega_1}{2}\left(1-\frac{D}{\omega_I}S_{2z}\right)\left(S_{1+}+S_{1-}\right)+}\\[3mm]
{\displaystyle+\left(\frac{D^2}{4\omega_I}+2DS_{2z}+\sum_kA_{1k}I_{kz}\right)S_{1z}+}\\[4mm]
{\displaystyle+\left(\frac{2\omega_1^2-D^2}{4\omega_I}+\sum_kA_{2k}I_{kz}\right)S_{2z}+}\\[4mm]
{\displaystyle+\sum_k\Big[\left(b_{k0}+b_{k1}S_{1z}S_{2z}\right)I_{kz}-}\\[2mm]
{\displaystyle-\frac{\omega_1}{4\omega_I}\left(B_{2k}Y_k+B_{2k}^*Y_k^\dagger\right)-
\frac{D}{4\omega_I}\left(\bar B_kZ_k+\bar B_k^*Z_k^\dagger\right)\Big],}\\[4mm]
{\displaystyle{\mathcal D}_0=\sum_{j=1}^2\left[\bar\Gamma^S_{j+}{\mathcal L}(S_{j+})+\bar\Gamma^S_{j-}{\mathcal L}(S_{j-})\right]+}\\[2mm]
{\displaystyle+\sum_{k=1}^N\Gamma^I_k\left[{\mathcal L}(I_{k+})+{\mathcal L}(I_{k-})\right]+
\Gamma^{SS}\left[{\mathcal L}(K)+{\mathcal L}(K^\dagger)\right]}
\end{array}
$$
with
$$
\begin{array}{c}
{\displaystyle b_{k0}=\frac{\vert B_{1k}\vert^2+\vert B_{2k}\vert^2}{8\omega_I},\
b_{k1}=\frac{B_{1k}B_{2k}^*+B_{1k}^*B_{2k}}{2\omega_I},}\\[4mm]
\bar B_k=B_{1k}-B_{2k},\
Y_k=I_{k+}S_{2-},\
Z_k=I_{k+}S_{1+}S_{2-},\\[2mm]
{\displaystyle\bar\Gamma_{1\pm}^S=\frac{R_1(1\mp p)}{2},\
\bar\Gamma_{2\pm}^S=\frac{R_1(1\mp p)}{2}+\frac{\omega_1^2R_2}{2\omega_I^2},}\\[4mm]
{\displaystyle\Gamma_k^I=\frac{r_1}{2}+\frac{(b_{k0}+b_{k1}S_{1z}S_{2z})(r_2+R_1)}{\omega_I},}\\[4mm]
{\displaystyle\Gamma^{SS}=\frac{D^2R_2}{\omega_I^2},\
K=S_{1+}S_{2-},\
X_{kk'}=I_{k+}I_{k'-}.}
\end{array}
$$

\subsection{Elimination of non-Zeeman orders}

The next and final step of the procedure is to eliminate the non-Zeeman orders $S_{1\pm}$, $Y_k,\,Y_k^\dagger$, $Z_k,\,Z_k^\dagger$ and $X_{kk'},\,X_{kk'}^\dagger$. These orders induce up and down quantum jumps between relevant Zeeman states of the spin system. Hence we can use the method described in the previous section, i.e., formula (\ref{tr}) for a fictitious spin-1/2 $\mathbf S$ chosen accordingly to the quantum jump in question. 

Precisely, the operators $S_{1\pm}$ realize jumps between the up and down orientations of the first electron, so in notations of the previous section
$$
\begin{array}{c}
\bar R_2=R_2,\
{\displaystyle\bar\omega_1=\omega_1\left(1-\frac{D}{\omega_I}S_{2z}\right),}\\[2mm]
{\displaystyle\bar\Delta=P_1\equiv\frac{D^2}{4\omega_I}+2DS_{2z}+\sum_k\left(A_{1k}+b_{k1}S_{2z}\right)I_{kz},}
\end{array}
$$   
characterizing respectively the effective transversal relaxation rate, Rabi frequency and detuning (the energy gap between the states in question). The operators $Y_k,\,Y_k^\dagger$ realize jumps between the up-down and down-up orientations of the second electron and the $k$th nucleus leading to
$$
\begin{array}{c}
{\displaystyle\bar R_2=R_2,\
\bar\omega_1=\Big\vert\frac{\omega_1B_{2k}}{2\omega_I}\Big\vert,}\\[3mm]
\bar\Delta=P_{2k}\equiv b_{k0}+\left(A_{1k}-2D\right)S_{1z}+\\[2mm]
{\displaystyle+\sum_{k'\not=k}\left(2d_{kk'}-A_{2k'}-b_{k'1}S_{1z}\right)I_{k'z}.}
\end{array}
$$ 
The operators $Z_k,\,Z_k^\dagger$ realize jumps between the up-down-up and down-up-down orientations of the electrons and the $k$th nucleus leading to
$$
\begin{array}{c}
{\displaystyle\bar R_2=2R_2,\
\bar\omega_1=\Big\vert\frac{D\bar B_k}{2\omega_I}\Big\vert,}\\[4mm]
{\displaystyle\bar\Delta=P_{3k}\equiv b_{k0}+\frac{D^2-\omega_1^2}{2\omega_I}+}\\[3mm]
{\displaystyle+\sum_{k'\not=k}\left(2d_{kk'}-A_{2k'}+A_{1k'}\right)I_{k'z}.}
\end{array}
$$ 
The operators $X_{kk'},\,X_{kk'}^\dagger$ realize jumps between the up-down and down-up orientations of the $k$th and $k'$th nuclei leading to
$$
\begin{array}{c}
\bar R_2=2r_2,\
\bar\omega_1=d_{kk'},\\[3mm]
\bar\Delta=\bar P_{kk'}\equiv b_{k0}-b_{k'0}+\left(b_{k1}-b_{k'1}\right)S_{1z}S_{2z}+\\[3mm]
{\displaystyle+\sum_k\left(A_{jk}-A_{jk'}\right)S_{jz}+2\sum_{s\not=k,k'}\left(d_{ks}-d_{k's}\right).}
\end{array}
$$ 
Here we used the condition $R_2\gg R_1,\,r_{1,2}$ valid in low-temperature DNP experiments. 

Using formula (\ref{tr}) with the notation $\eta'=(\bar R_2^2+\bar\Delta^2)/\bar\omega_1^2$, we come to the final master equation $\dot\rho_Z=\bar{\mathcal D}\rho_Z$ closed in the Zeeman subspace, with the right-hand side given by the formula
$$
\begin{array}{c}
\bar{\mathcal D}={\mathcal D_0}+\bar\Gamma^S\left[{\mathcal L}(S_{1+})+{\mathcal L}(S_{1-})\right]+\\[1mm]
{\displaystyle\sum_{k=1}^N\Gamma^{SI}_k\left[{\mathcal L}(Y_k)+{\mathcal L}(Y_k^\dagger)\right]+}\\[5mm]
{\displaystyle+\sum_{k<k'}\Gamma^{II}_{kk'}\left[{\mathcal L}(X_{kk'})+{\mathcal L}(X_{kk'}^\dagger)\right]+}\\[4mm]
{\displaystyle+\sum_{k=1}^N\Gamma^{SSI}_k\left[{\mathcal L}(Z_k)+{\mathcal L}(Z_k^\dagger)\right]}
\end{array}
$$
where
$$
\begin{array}{c}
{\displaystyle\bar\Gamma^S=\frac{\omega_1^2}{2}R_2\left(1-\frac{D}{\omega_I}S_{2z}\right)^2\left(R_2^2+P_1^2\right)^{-1},}\\[4mm]
{\displaystyle\Gamma_k^{SI}=\frac{1}{2}\,\Big\vert\frac{\omega_1B_{2k}}{2\omega_I}\Big\vert^2R_2\left(R_2^2+P_{2k}^2\right)^{-1},}\\[4mm]
{\displaystyle\Gamma_k^{SSI}=\Big\vert\frac{D\bar B_k}{2\omega_I}\Big\vert^2R_2\left(4R_2^2+P_{3k}^2\right)^{-1},}\\[3mm]
\Gamma_{kk'}^{II}=d_{kk'}^2r_2\left(4r_2^2+\bar P_{kk'}^2\right)^{-1}.
\end{array}
$$
Noting that in the Zeeman subspace
$$
{\mathcal L}(\Xi)+{\mathcal L}(\Xi^\dagger)={\mathcal L}(\Xi+\Xi^\dagger),
$$ 
we come to the formulas for the dissipators ${\mathcal D}_{1-3}$ given in the main text.

\noindent
{\bf\em Acknowledgments} --- The research leading to these results has received funding from the European Research Council under the European Union's Seventh Framework Programme (FP/2007-2013) / ERC Grant Agreement No. 335266 (ESCQUMA) and the EPSRC Grant No. EP/N03404X/1.

\bibliography{CEDNP_Lindblad}{}

\end{document}